\documentclass[a4paper, oneside, 11pt]{article}%
\usepackage{amsmath}
\usepackage{amsfonts}
\usepackage{amssymb}
\usepackage{graphicx}
\usepackage{appendix}
\usepackage[square, numbers, sort&compress]{natbib}
\usepackage{color}%
\providecommand{\U}[1]{\protect\rule{.1in}{.1in}}

\newtheorem{theorem}{Theorem}[section]

\newtheorem{corollary}[theorem]{Corollary}

\newtheorem{definition}[theorem]{Definition}

\newtheorem{lemma}[theorem]{Lemma}

\newtheorem{remark}[theorem]{Remark}

\numberwithin{equation}{section}

\newcommand{\E}{{\mathbb E}}
\newcommand{\R}{{\mathbb R}}
\newcommand{\pf}{\noindent\textbf{Proof:} }
\newcommand{\eof}{\hfill{$\Box$}}

\usepackage{url}
\usepackage{hyperref}
\hypersetup{
colorlinks=true, 
breaklinks=true, 
urlcolor=green , 
linkcolor=red, 
citecolor=blue, 
pdftitle={}, 
pdfauthor={}, 
pdfsubject={} 
}

\newcommand{\dist}{\ensuremath{\mathrm{dist}}}

\usepackage[normalem]{ulem}

\renewcommand{\geq}{\geqslant}
\renewcommand{\leq}{\leqslant}
\newcommand{\nn}{\nonumber}

\begin{document}

\title{Constrained monotone mean-variance problem with random coefficients}
\author{Ying Hu \thanks{Univ Rennes, CNRS, IRMAR-UMR 6625, F-35000 Rennes, France. Partially supported by Lebesgue
Center of Mathematics \textquotedblleft Investissements d'avenir\textquotedblright program-ANR-11-LABX-0020-01. Email:
\texttt{ying.hu@univ-rennes1.fr }}
\and Xiaomin Shi\thanks{School of Statistics and Mathematics, Shandong University of Finance and Economics, Jinan
250100, China. Partially supported by NSFC (No.~11801315), NSF of Shandong Province (No.~ZR2018QA001). Email: \texttt{shixm@mail.sdu.edu.cn}}
\and Zuo Quan Xu\thanks{Department of Applied Mathematics, The Hong Kong Polytechnic University, Kowloon, Hong Kong.
Partially supported by NSFC (No.~11971409), Hong Kong
RGC (GRF ~15202421 and 15204622), the PolyU-SDU Joint Research Center on Financial Mathematics, the CAS AMSS-PolyU Joint Laboratory of Applied Mathematics, the Research Centre for Quantitative Finance (1-CE03), and the Hong Kong Polytechnic University. Email: \texttt{maxu@polyu.edu.hk}}}
\maketitle
This paper studies the monotone mean-variance (MMV) problem and the classical mean-variance (MV) problem with convex cone trading constraints in a market with random coefficients. We provide semiclosed optimal strategies and optimal values for both problems via certain backward stochastic differential equations (BSDEs). After noting the links between these BSDEs, we find that the two problems share the same optimal portfolio and optimal value. This generalizes the result of 
Shen and Zou $[$  SIAM J. Financial Math., 13 (2022), pp. SC99-SC112$]$
from deterministic coefficients to random ones.

\smallskip
{\textbf{Key words}.} monotone mean-variance, cone constraints, random coefficients, robust control
\smallskip
\textbf{Mathematics Subject Classification (2020)} 91B16 93E20 60H30 91G10

\addcontentsline{toc}{section}{\hspace*{1.8em}Abstract}

\section{Introduction}
The mean-variance (MV) portfolio selection theory, pioneered by Markowitz \cite{Mark} in a single-period setting, has become a cornerstone in modern finance theory. Zhou and Li \cite{ZL} study the MV problem in a continuous-time setting using the embedding technique and the well-developed stochastic linear-quadratic (LQ) control theory. Although the MV model is a Nobel-Prize-winning work, the feature of its non-monotonicity brings troublesome in development because monotonicity is one of most compelling principles of economic rationality. To overcome the major drawback, Maccheroni et al. \cite{MMRT} propose and solve, in a single-period setting, the monotone mean-variance (MMV) model in which the objective functional is the best approximation of the MV functional among those which are monotone. \u{C}ern\'y et al. \cite{CMMR} connect the MMV portfolio selection problem to optimization of truncated quadratic utility.
Trybu{\l}a and Zawisza \cite{TZ} consider the MMV problem in a stochastic factor model, hence an incomplete market. By applying the
Hamilton-Jacobi-Bellman-Isaacs (HJBI) equations approach, they obtain the optimal investment strategy and the optimal value which coincide with that for the classical MV problem. Later, when the underlying asset prices are continuous, Strub and Li \cite{SL} prove that the terminal wealth levels corresponding to the optimal portfolio strategies for both the MMV and the MV problems will drop in the domain of the monotonicity of the classical MV functional. This, with earlier results in \cite{MMRT}, leads to the conclusion that the optimal strategies for the MMV and the MV problems always coincide when the portfolios are unconstrained and the underlying asset prices are continuous. With conic convex constraints on the portfolios, Shen and Zou \cite{SZ} solve the MMV and the MV problems successively by means of the HJBI equation approach, and find that the optimal strategies to both problems coincide.
Please refer to \cite{SZ} for a more detailed review of the MMV problems and their links with the classical MV problems, and to \u{C}ern\'y \cite{Cerny} for a new characterization of the MMV model.

This paper aims to generalize \cite{SZ} to a diffusion model with random coefficients (including excess return rates and volatility rate) and cone trading constraints. We follow the same direction as \cite{SZ} and \cite{TZ}, i.e. first solving the MMV and the MV problems, and then comparing their optimal portfolios and optimal values to see whether they coincide. A strong motivation for taking this direction is the statement ``It is certainly possible to consider problems in a more general setting, but explicit solutions are likely unavailable..." in \cite[the footnote of page SC101]{SZ}. In fact, the MMV problem is a robust control (optimization) problem which barely has an explicit solution compared with the MV problem. Fortunately, we firstly guess an optimal portfolio candidate, via a heuristic argument, by means of a specific backward stochastic differential equations (BSDE) and then prove a verification theorem rigorously. Along this line, we eventually provide a semiclosed solution for the constrained MMV problem with random coefficients. This is the main mathematical contribution of this paper. As for the constrained MV problem with random coefficients, Hu and Zhou \cite{HZ} give the solution in a slightly different setting of minimizing a portfolio's variance subject to the constraint that its expected return equals a prescribed level. Therefore, in this paper, we need to generalize a little bit from some results in \cite{HZ} to solve the present MV problem besides some tedious calculations to get the extreme values of certain deterministic quadratic functions. In the end, after noticing the links between two related BSDEs, the coincidence of the optimal portfolios and optimal values in the MMV and the MV problems is achieved.

The rest part of this paper is organized as follows.
In Section \ref{fm}, we present the financial market and formulate the constrained MMV problem with random coefficients. In Section \ref{Heuri}, we derive heuristically the BSDEs and the optimal candidate. Section \ref{Verification} provides a rigorous verification for the optimal strategies and optimal value. In Section \ref{Comparison}, we solve the constrained MV problem with random coefficients and make a comparison with the MMV problem. Some concluding remarks are given in Section \ref{conclude}.

\section{Problem formulation}\label{fm}
Let $(\Omega, \mathcal F, \mathbb{P})$ be a fixed complete probability space on which is defined a standard $n$-dimensional Brownian motion $W_t=(W_{1, t}, \ldots, W_{n, t})'$.
Define the filtration $\mathcal F_t=\sigma\{W_s: 0\leq s\leq t\}\bigvee\mathcal{N}$, where $\mathcal{N}$ is the totality of all the $\mathbb{P}$-null sets of $\mathcal{F}$.

%
Let $\R^m$ be the set of $m$-dimensional column vectors, let $\R^m_+$ be its subset of vectors whose components are nonnegative, and let $\R^{m\times n}$ be the set of $m\times n$ real matrices. We denote the transpose of $M$ by $M'$, and the norm by $|M|=\sqrt{\textrm{trace}(MM')}$.
Let $\mathbb{S}^n$ be the set of symmetric $n\times n$ real matrices.
We write $M>$ (resp., $\geq$) $0$ for any positive definite (resp., positive semidefinite) matrix $M\in\mathbb{S}^n$. We write $A>$ ($\geq$) $B$ if $A, B\in\mathbb{S}^n$ and $A-B>$ ($\geq$) $0.$

Define the following spaces:
\begin{align*}
L^{0}_{\mathcal F}(0, T;\mathbb{R})&=\Big\{\varphi:[0, T]\times\Omega\rightarrow
\mathbb{R}\;\Big|\; \varphi\mbox{ is an }\{\mathcal{F}%
_{t}\}_{t\geq0}\mbox{-predictable process}\Big\}, \\
L^{2}_{\mathcal F}(0, T;\mathbb{R})&=\Big\{\varphi\in L^{0}_{\mathcal F}(0, T;\mathbb{R})\;\Big|\;\E\int_{0}^{T}|\varphi_t|^{2}dt<\infty
\Big\}, \\
L^{\infty}_{\mathcal{F}}(0, T;\mathbb{R})&=\Big\{\varphi\in L^{0}_{\mathcal F}(0, T;\mathbb{R})\;\Big|\; \varphi\mbox{ is essentially bounded} \Big\}.
\end{align*}

Consider a financial market consisting of a risk-free asset (the money market
instrument or bond) whose price is $S_{0}$ and $m$ risky securities (the
stocks) whose prices are $S_{1}, \ldots, S_{m}$. And assume $m\leq n$, i.e., the number of risky securities is no more than the dimension of the Brownian motion.
The asset prices $S_k$, $k=0, 1, \ldots, m, $ are driven by SDEs:
\begin{align*}
\begin{cases}
dS_{0,t}=r_tS_{0,t}dt, \\
S_{0,0}=s_0,
\end{cases}
\end{align*}
and
\begin{align*}
\begin{cases}
dS_{k,t}=S_{k,t}\Big((\mu_{k,t}+r_t)dt+\sum\limits_{j=1}^n\sigma_{kj,t}dW_{j, t}\Big), \\
S_{k,0}=s_k,
\end{cases}
\end{align*}
where, for every $k=1, \ldots, m$, $r$ is the interest rate process and $\mu_k$ and $\sigma_k:=(\sigma_{k1}, \ldots, \sigma_{kn})$ are the mean excess return rate process and volatility rate process of the $k$th risky security.

Define the mean excess return vector
$\mu=(\mu_1, \ldots, \mu_m)'$
and volatility matrix
\begin{align*}
\sigma=
\left(
\begin{array}{c}
\sigma_1\\
\vdots\\
\sigma_m\\
\end{array}
\right)
\equiv (\sigma_{kj})_{m\times n}.
\end{align*}
Throughout this paper, we assume the interest rate $r$ is a bounded deterministic function of $t$,
\begin{align*}
\mu\in L_{\mathcal{F}}^\infty(0, T;\mathbb{R}^m), \
\sigma\in L_{\mathcal{F}}^\infty(0, T;\mathbb{R}^{m\times n}).
\end{align*}
Also, there exists a constant $\delta>0$ such that
$\sigma\sigma'\geq \delta I_m$ for a.e. $t\in[0, T]$, where $I_m$ denotes the $m$-dimensional identity matrix. In particular,
$\sigma\phi=\mu$, where
\begin{align}\label{pk}
\phi=\sigma'(\sigma\sigma')^{-1}\mu
\end{align}
is called the pricing kernel of the market.

Consider a small investor (``He'') whose actions cannot affect the asset prices. He will decide at every time
$t\in[0, T]$ the amount $\pi_{j,t}$ of his wealth to invest in the $j$th risky asset, $j=1, \ldots, m$. The vector process $\pi:=(\pi_1, \ldots, \pi_m)'$ is called a portfolio of the investor. Then the investor's self-financing wealth process $X$ corresponding to a portfolio $\pi$ is the unique strong solution of the SDE:
\begin{align}
\label{wealth}
\begin{cases}
dX_t=(r_tX_t+\pi_t'\mu_t)dt+\pi_t'\sigma_tdW_t, \\
X_0=x.
\end{cases}
\end{align}

Let $\Gamma\subseteq\mathbb{R}^m$ be a given closed convex cone; i.e., $\Gamma$ is closed  and convex, and if $u\in\Gamma$, then $\lambda u\in\Gamma$ for all real $\lambda\geq 0$.\footnote{The cone constraint in \cite[section 5]{HSX} shall also be convex cone constraint, for otherwise the dual method may not work. } It stands for the constraint set for portfolios.
The class of admissible portfolios is defined as the set
\begin{align*}
\Pi:=\Big\{\pi\in L^2_\mathcal{F}(0, T;\mathbb{R}^m)\;\Big|\; \pi \in\Gamma, \mbox{ a.e. a.s.}\Big\}.
\end{align*}
And we will denote by $X^{\pi}$ the wealth process \eqref{wealth} whenever it is necessary to indicate its dependence on $\pi\in\Pi$.

For $(t,\omega)\in[0,T]\times\Omega$, define the set $\sigma_t(\omega)'\Gamma\subseteq\mathbb{R}^n$ by
\[
\sigma_t(\omega)'\Gamma=\big\{\sigma_t(\omega)'\pi\;\big|\;\pi\in\Gamma\big\}.
\]
For every $(t,\omega)$, the set $\sigma_t(\omega)'\Gamma$ is a closed convex cone.

For any process $\eta\in L^{0}_{\mathcal{F}}(0,T;\R^n)$ such that
\begin{align}\label{GirLam}
\Lambda^\eta_t:=\mathcal{E}(\int_0^t\eta'dW)
\end{align}
is a martingale, define $\mathbb{P}^{\eta}$ by
\begin{align}\label{GirP}
\frac{d\mathbb{P}^{\eta}}{d\mathbb{P}}\bigg|_{\mathcal{F}_t}=\Lambda^\eta_t.
\end{align}
By Girsanov's theorem,
\begin{align}\label{GirW}
W^\eta_t:=W_t-\int_0^t\eta_sds
\end{align}
is a Brownian motion under $\mathbb{P}^{\eta}$.
Denote
\begin{align*}
\mathcal{A}:=\Big\{\eta\in L^0_\mathcal{F}(0, T;\mathbb{R}^m)\;\Big|\; \E[\Lambda^\eta_t]=1 \mbox{ and } \E[(\Lambda^\eta_t)^2]<\infty \mbox{ for all } t\in[0,T]\Big\}.
\end{align*}
In this paper, we first solve the following MMV problem:
\begin{align}\label{MMV}
\sup_{\pi\in\Pi}\inf_{\eta\in\mathcal{A}}\E^{\mathbb{P}^{\eta}}\Big[X_T+\frac{1}{2\theta}(\Lambda^\eta_T-1)\Big],
\end{align}
where $\theta$ is a given positive constant.

\section{
Optimal candidate: a heuristic derivation}
\label{Heuri}
In order to solve the problem \eqref{MMV}, we hope, via a heuristic argument, to find a family of stochastic process $R^{(\eta,\pi)}$ and a pair $(\hat\eta,\hat\pi)$ with the following properties:
\begin{enumerate}
\item $R^{(\eta,\pi)}_T=X_T+\frac{1}{2\theta}(\Lambda^\eta_T-1)$ for all $(\eta,\pi)\in\mathcal{A}\times\Pi$.

\item $R^{(\eta,\pi)}_0=R_0$ is constant for all $(\eta,\pi)\in\mathcal{A}\times\Pi$.

\item $\E^{\mathbb{P}^{\hat\eta}}\Big[X_T^{\pi}+\frac{1}{2\theta}(\Lambda^{\hat\eta}_T-1)\Big]
\leq R_0$ for all $\pi\in\Pi$.

\item $\E^{\mathbb{P}^{\eta}}\Big[X_T^{\hat\pi}+\frac{1}{2\theta}(\Lambda^{\eta}_T-1)\Big]\geq R_0$ for all $\eta\in\mathcal{A}$.

\item $\E^{\mathbb{P}^{\hat\eta}}\Big[X_T^{\hat\pi}+\frac{1}{2\theta}(\Lambda^{\hat\eta}_T-1)\Big]=R_0$.
\end{enumerate}
Once this is done, we will then rigorously show that $(\hat\eta,\hat\pi)$ is an optimal solution for the problem \eqref{MMV} and $R_0$ is its optimal value.

To derive the above family, we let
\[
R^{(\eta,\pi)}_t=X_t^{\pi}h_t+\frac{1}{2\theta}(\Lambda_t^{\eta} Y_t-1), \ t\in[0,T], \ (\eta,\pi)\in\mathcal{A}\times\Pi,
\]
where $\Lambda^\eta_{t}$ is defined by \eqref{GirLam};
$(Y,Z)$ satisfies some BSDE
\begin{align}\label{Y0}
\begin{cases}
dY=-fdt+Z_t'dW,\\
Y_T=1,\quad Y>0,
\end{cases}
\end{align}
and $(h,L)$ satisfies some BSDE
\begin{align}\label{H}
\begin{cases}
dh=-gdt+L_t'dW,\\
h_T=1.
\end{cases}
\end{align}
Our aim reduces to finding proper drivers $f$ and $g$, which are independent of $(\eta,\pi)$, for the above two BSDEs.
Notice the first property is already satisfied by the choice of boundary conditions given above.

Recalling \eqref{GirW}, in terms of $W^\eta$, we have
\begin{align*}
\begin{cases}
dY=(-f+Z'\eta)dt+Z'dW^\eta,\\
Y_T=1,\quad Y>0,
\end{cases}
\end{align*}
and
\begin{align*}
\begin{cases}
dh=(-g+L'\eta)dt+L'dW^\eta,\\
h_T=1.
\end{cases}
\end{align*}
Applying
It\^{o}'s formula to $X_th_t$ and $\Lambda_t Y_t$, we have
\begin{align*}
d(Xh)=[rhX-gX+L'\eta X+\pi'(h\mu+h\sigma\eta+\sigma L)]dt+(\cdots)dW^{\eta},
\end{align*}
and
\begin{align*}
d(\Lambda Y)=[\Lambda Y|\eta|^2+2\Lambda\eta' Z-\Lambda f]dt+(\cdots)dW^{\eta}.
\end{align*}
Combining the above two equations, we obtain
\begin{align*}
dR^{(\eta,\pi)}&=\Big[\frac{1}{2\theta}\Lambda Y|\eta|^2+\frac{1}{\theta}\Lambda \eta'Z+h\eta'\sigma'\pi+X\eta' L+(rh-g)X+\pi'(h\mu+\sigma L)-\frac{1}{2\theta}\Lambda f\Big]dt\\
&\quad\;+(\cdots)dW^{\eta}.
\end{align*}
Taking $g:=rh$, and noting that $r$ is a deterministic function of $t$, we obtain the solution of \eqref{H}:
\begin{align}\label{h}
h_t=e^{\int_t^Tr_sds}, \ L_t=0, \ t\in[0,T].
\end{align}
This in particular implies that
\[
R^{(\eta,\pi)}_0=X_0^{\pi}h_0+\frac{1}{2\theta}(\Lambda_0^{\eta} Y_0-1)
=xh_0+\frac{1}{2\theta}(Y_0-1):=R_0
\]
is a constant for all $(\eta,\pi)$. So the second property follows.

Since $Y>0$,
\begin{align*}
dR^{(\eta,\pi)}&=\Big[\frac{1}{2\theta}\Lambda Y|\eta|^2+\frac{1}{\theta}\Lambda \eta'Z+h\eta'\sigma'\pi+h\pi'\mu-\frac{1}{2\theta}\Lambda f\Big]dt+(\cdots)dW^{\eta}\\
&=\Big[\frac{\Lambda Y}{2\theta}\Big|\eta+\frac{\theta}{\Lambda Y}\Big(\frac{1}{\theta}\Lambda Z+h\sigma'\pi\Big)\Big|^2-\frac{\Lambda Y}{2\theta}\frac{\theta^2}{\Lambda^2Y^2}\Big|\frac{1}{\theta}\Lambda Z+h\sigma'\pi\Big|^2+h\pi'\mu-\frac{1}{2\theta}\Lambda f\Big]dt\\
&\quad\;+(\cdots)dW^{\eta}.
\end{align*}
Integrating from $0$ to $T$ and taking expectation $\E^{\mathbb{P}^{\eta}}$, we have
\begin{align}
\E^{\mathbb{P}^{\eta}}[R^{(\eta,\pi)}_{T}]=R_0
+\E^{\mathbb{P}^{\eta}}\int_0^{T} \frac{\Lambda }{2\theta}\bigg\{ & Y\Big|\eta
+\frac{1}{ Y}\Big(Z+\frac{\theta}{\Lambda} h\sigma'\pi\Big)\Big|^2\nonumber\\
&-\Big[f+\frac{1}{Y}\Big| Z+\frac{\theta}{\Lambda} h\sigma'\pi\Big|^2-\frac{2\theta}{\Lambda} h\pi'\sigma\phi\Big]\bigg\}ds \label{exp}
\end{align}
on recalling that $\phi$ is the pricing kernel, defined by \eqref{pk}.

To fulfill the third, fourth and fifth properties,
since $\theta, \Lambda>0$, {it suffices to} find $f$, $\hat\pi$, and $\hat\eta$ such that
\begin{align} \label{ineq2}
Y\Big|\hat\eta+\frac{1}{ Y}\Big(Z+\frac{\theta}{\Lambda} h\sigma'\pi\Big)\Big|^2-\Big[f+\frac{1}{Y}\Big| Z+\frac{\theta}{\Lambda} h\sigma'\pi\Big|^2-\frac{2\theta}{\Lambda} h\pi'\sigma\phi\Big]\leq 0
\end{align}
for all $\pi\in\Pi$;
\begin{align*} 
Y\Big|\eta+\frac{1}{ Y}\Big(Z+\frac{\theta}{\Lambda} h\sigma'\hat\pi\Big)\Big|^2-\Big[f+\frac{1}{Y}\Big| Z+\frac{\theta}{\Lambda} h\sigma'\hat\pi\Big|^2-\frac{2\theta}{\Lambda} h\hat\pi'\sigma\phi\Big] \geq 0
\end{align*}
for all $\eta\in\mathcal{A}$;
and
\begin{align*} 
Y\Big|\hat\eta+\frac{1}{ Y}\Big(Z+\frac{\theta}{\Lambda} h\sigma'\hat\pi\Big)\Big|^2-\Big[f+\frac{1}{Y}\Big| Z+\frac{\theta}{\Lambda} h\sigma'\hat\pi\Big|^2-\frac{2\theta}{\Lambda} h\hat\pi'\sigma\phi\Big]= 0.
\end{align*}
These motivate us to conjecture the following:
\begin{align}
\hat\eta=-\frac{1}{ Y}\Big(Z+\frac{\theta}{\Lambda^{\hat\eta}} h\sigma'\hat\pi\Big),
\end{align}
and
\begin{align}
f=-\frac{1}{Y}\Big| Z+\frac{\theta}{\Lambda^{\hat\eta}} h\sigma'\hat\pi\Big|^2+\frac{2\theta}{\Lambda^{\hat\eta}} h\hat\pi'\sigma\phi.
\end{align}
To find a proper $\hat\pi$, taking the above two expressions into \eqref {ineq2}, we get
\begin{align}
&\quad\; Y\Big|-\frac{1}{ Y}\Big(Z+\frac{\theta}{\Lambda} h\sigma'\hat\pi\Big)+\frac{1}{ Y}\Big(Z+\frac{\theta}{\Lambda} h\sigma'\pi\Big)\Big|^2\nonumber\\
& \leq -\frac{1}{Y}\Big| Z+\frac{\theta}{\Lambda} h\sigma'\hat\pi\Big|^2+\frac{2\theta}{\Lambda} h\hat\pi'\sigma\phi+\frac{1}{Y}\Big| Z+\frac{\theta}{\Lambda} h\sigma'\pi\Big|^2-\frac{2\theta}{\Lambda} h\pi'\sigma\phi.
\end{align}
Write $u=\frac{\theta}{\Lambda} h\sigma'\pi$ and $\xi=\frac{\theta}{\Lambda} h\sigma'\hat\pi$; then the above reads
\begin{align*}
\big|u-\xi\big|^2 +\big| Z+\xi\big|^2 -\big| Z+u\big|^2
-2Y\phi'\xi+2Y\phi' u\leq 0,
\end{align*}
namely,
\begin{align} \label{projcon}
(Y\phi-Z-\xi)'(u-\xi)\leq 0.
\end{align}
Because $\theta$, $\Lambda$, $h>0$ and $\Gamma$ is a cone, we see $\pi\in \Gamma$ if and only if $u\in \sigma'\Gamma $.

The following lemma is the second projection theorem; see, e.g., \cite[Theorem 9.8]{Beck}.
\begin{lemma}\label{proj}
Suppose $\xi\in \sigma'\Gamma$.
Then the inequality \eqref{projcon} holds for all $u\in \sigma'\Gamma $ if and only if
\begin{align} \label{xi}
\xi=\mbox{Proj}_{\sigma'\Gamma}(\phi Y-Z).
\end{align}
Here, $\mbox{Proj}_{C}(a)$ denotes the projection of $a$ to a closed convex set
$C$, which is uniquely determined by
\[|a-\mbox{Proj}_{C}(a)|=\min_{b\in C}|a-b|.\]
\end{lemma}
Consequently,
\begin{align}
f &=-\frac{1}{Y}\Big| Z+\frac{\theta}{\Lambda} h\sigma'\hat\pi\Big|^2+\frac{2\theta}{\Lambda} h\hat\pi'\sigma\phi\label{funf2}\\
&=-\frac{1}{Y}\Big| Z+\mbox{Proj}_{\sigma'\Gamma}(Y\phi-Z)\Big|^2+2\phi'\mbox{Proj}_{\sigma'\Gamma}(Y\phi-Z)\nonumber.
\end{align}
Overall, we conjecture that
\begin{align}
f&=-\frac{1}{Y}\Big| Z+\mbox{Proj}_{\sigma'\Gamma}(Y\phi-Z)\Big|^2+2\phi'\mbox{Proj}_{\sigma'\Gamma}(Y\phi-Z),\label{funf}\\
\hat\pi &=\frac{\Lambda^{\hat\eta}} {h\theta}(\sigma\sigma')^{-1} \sigma\mbox{Proj}_{\sigma'\Gamma}(Y\phi-Z), \label{hatpi}\\
\hat\eta &=-\frac{1}{ Y}\Big(Z+\mbox{Proj}_{\sigma'\Gamma}(Y\phi-Z)\Big).\label{saddleeta}
\end{align}

Let $f$ be defined by \eqref{funf} and $\xi$ by \eqref{xi}; then
\begin{align}
f&=-\frac{1}{Y}\big| Z+\xi\big|^2+2\phi'\xi\notag\\
&=-\frac{1}{Y}|\xi|^2+\frac{2}{Y}(\phi Y-Z)'\xi -\frac{1}{Y}|Z|^2\notag\\
&=-\frac{1}{Y} \Big|\xi-(\phi Y- Z)\Big|^2+\frac{1}{Y}|\phi Y- Z|^2-\frac{1}{ Y}|Z|^2\notag\\
&=-\frac{1}{Y}\inf_{ \pi\in\Gamma}\Big|\sigma' \pi-(\phi Y- Z)\Big|^2
+\frac{1}{Y}|\phi Y- Z|^2-\frac{1}{ Y}|Z|^2\notag\\
&=-\frac{1}{Y}\inf_{ \pi\in\Gamma}\Big[\pi'\sigma\sigma' \pi-2\pi'\sigma(\phi Y- Z)\Big]-\frac{1}{ Y}|Z|^2\label{fexp}
\end{align}
So the desired BSDE \eqref{Y0} is
\begin{align}\label{Y}
\begin{cases}
dY=\Big[\frac{1}{Y}\inf_{ \pi\in\Gamma}\Big[\pi'\sigma\sigma' \pi-2\pi'\sigma(\phi Y- Z)\Big]+\frac{1}{ Y}|Z|^2\Big]dt+Z'dW,\\
Y_T=1,\quad Y>0.
\end{cases}
\end{align}

\section{Solution}\label{Verification}

\begin{definition}\label{def}
A pair $(Y,Z)$ is called a solution to the BSDE \eqref{Y} if it satisfies \eqref{Y} and $(Y, Z)\in L^\infty_{\mathcal{F}^W}(0, T; \mathbb{R})\times L^{2}_{\mathcal{F}}(0,T;\mathbb{R}^m)$. The solution is called uniformly positive if $Y\geq \delta$ for all $t\in[0, T]$, a.s. with some deterministic constant $\delta>0$.
\end{definition}

\begin{theorem}
There exists a unique uniformly positive solution $(Y,Z)$ to the BSDE \eqref{Y}.
\end{theorem}
\pf
Consider
\begin{align}
\label{P}
\begin{cases}
dP
=-\inf\limits_{\pi\in\Gamma}\Big[P\pi'\sigma\sigma'\pi-2\pi'(P\mu+\sigma\Delta)\Big]dt+\Delta'dW, \\
P_T=1, \quad P_t>0.
\end{cases}
\end{align}
From \cite[Theorem 4.2 and Theorem 5.2]{HZ}, the above BSDE \eqref{P} admits a unique solution $(P,\Delta)\in L^{\infty}_{\mathcal{F}}(0, T;\mathbb{R})\times L^{2}_{\mathcal{F}}(0,T;\mathbb{R}^m)$ such that $c_1\leq P_t\leq c_2$ a.s. for all $t\in[0, T]$, with some deterministic constants $c_2>c_1>0$.
Then
\begin{align}\label{PtoY}
(Y,Z):=\Big(\frac{1}{P}, -\frac{\Delta}{P^2}\Big),
\end{align}
is well defined and $(Y,Z)\in L^{\infty}_{\mathcal{F}}(0,T;\mathbb{R})\times L^{2}_{\mathcal{F}}(0,T;\mathbb{R}^m)$ with $\frac{1}{c_2}\leq Y_t\leq \frac{1}{c_1}$ a.s. for all $t\in[0, T]$. It can be directly verified, using It\^{o}'s formula, that $(Y,Z)$ is a solution to \eqref{Y}.
The above change $(P,\Delta)\to (Y,Z)$ is invertible, so the uniqueness of \eqref{Y} follows from that of \eqref{P}.
\eof

\begin{lemma}
Let $(Y,Z)$ be the unique uniformly positive solution to \eqref{Y}.
Let $h$, $\hat\pi, \hat\eta$ be defined in \eqref{h}, \eqref{hatpi}, and \eqref{saddleeta}, respectively.
Then
\begin{align}\label{LambdatoX}
\theta h_tX_t^{\hat\pi}+Y_t\Lambda_t^{\hat\eta}\equiv\theta h_0 x+Y_0, \ t\in[0,T].
\end{align}
\end{lemma}
\pf
Let $\xi=\mbox{Proj}_{\sigma'\Gamma}(\phi Y-Z)$ be defined by \eqref{xi}.
By taking $u$ as $2\xi$ and $0$ in \eqref{projcon}, respectively, we immediately have
\begin{align}\label{projpro}
-\xi'(\phi Y-Z-\xi)=|\xi|^2-\xi'(\phi Y-Z)=0.
\end{align}
Applying It\^{o}'s formula to $\theta h_tX_t^{\hat\pi}+Y_t\Lambda_t^{\hat\eta}$,
we have
\begin{align*}
d(\theta hX^{\hat\pi}+Y\Lambda^{\hat\eta})&=\Lambda^{\hat\eta}\xi'\phi dt+\Lambda^{\hat\eta}\xi'dW+\frac{\Lambda^{\hat\eta}}{Y}\Big[\dist_{\sigma'\Gamma}^2\Big(\phi Y-Z\Big)-|\phi Y-Z|^2-\xi'Z\Big]dt-\Lambda^{\hat\eta}\xi'dW\\
&=\frac{\Lambda^{\hat\eta}}{Y}\Big[\xi'\phi Y+|\xi-(\phi Y-Z)|^2-|\phi Y-Z|^2-\xi'Z\Big]dt\\
&=\frac{\Lambda^{\hat\eta}}{Y}\Big[\xi'\phi Y+|\xi|^2-2\xi'(\phi Y-Z)-\xi'Z \Big]dt\\
&=\frac{\Lambda^{\hat\eta}}{Y}\Big[|\xi|^2-\xi'(\phi Y-Z)\Big]dt=0.
\end{align*}
The claim thus follows.
\eof

\begin{theorem}[Verification]\label{verif}
Let $(Y,Z)$ be the unique uniformly positive solution to \eqref{Y}.
Let $h$, $\hat\pi, \hat\eta$ be defined in \eqref{h}, \eqref{hatpi}, and \eqref{saddleeta}, respectively.
Then
we have $\hat\pi\in\Pi$, $\hat\eta\in\mathcal{A}$, and
\begin{align}\label{saddle1}
\E^{\mathbb{P}^{\hat\eta}}\Big[X_T^{\pi}+\frac{1}{2\theta}(\Lambda^{\hat\eta}_T-1)\Big]
\leq xh_0+\frac{1}{2\theta}(Y_0-1), \ \forall \pi\in\Pi,
\end{align}
\begin{align}\label{saddle2}
\E^{\mathbb{P}^{\eta}}\Big[X_T^{\hat\pi}+\frac{1}{2\theta}(\Lambda^{\eta}_T-1)\Big]
\geq xh_0+\frac{1}{2\theta}(Y_0-1), \ \forall \eta\in\mathcal{A},
\end{align}
and
\begin{align}\label{saddle3}
\E^{\mathbb{P}^{\hat\eta}}\Big[X_T^{\hat\pi}+\frac{1}{2\theta}(\Lambda^{\hat\eta}_T-1)\Big]
=xh_0+\frac{1}{2\theta}(Y_0-1).
\end{align}
\end{theorem}
\pf
We divide the proof into three steps.

\textbf{Step 1:} To prove $\hat\pi\in\Pi$ and $\hat\eta\in\mathcal{A}$.
Let $\xi=\mbox{Proj}_{\sigma'\Gamma}(\phi Y-Z)$ be defined by \eqref{xi}; then there exists $\gamma\in\Gamma$ such that $\xi=\sigma'\gamma$. From \eqref{hatpi},
\begin{align*}
\hat\pi &=\frac{\Lambda} {h\theta}(\sigma\sigma')^{-1} \sigma\mbox{Proj}_{\sigma'\Gamma}(Y\phi-Z)=\frac{\Lambda} {h\theta}(\sigma\sigma')^{-1} \sigma \sigma'\gamma=\frac{\Lambda} {h\theta}\gamma,
\end{align*}
hence, by \eqref{LambdatoX},
\begin{align}\label{hatsigmapi2}
\sigma'\hat \pi=\frac{\Lambda} {h\theta}\sigma'\gamma&=\frac{\Lambda} {h\theta}\xi=\frac{\theta h_0x+Y_0-\theta h_t X^{\hat\pi}_t}{\theta h_tY_t}\xi=\frac{a- h_t X^{\hat\pi}_t}{ h_tY_t}\xi,
\end{align}
where $$a:=h_0x+Y_0/\theta.$$
Substituting \eqref{hatsigmapi2} into \eqref{wealth}, we have
\begin{align*}
\begin{cases}
dX_t^{\hat\pi}=(r_tX_t^{\hat\pi}+\frac{a- h_t X^{\hat\pi}_t}{ h_tY_t}\xi_t'\phi_t)dt+\frac{a- h_t X^{\hat\pi}_t}{ h_tY_t}\xi_t'dW_t, \\
X_0^{\hat\pi}=x.
\end{cases}
\end{align*}
Applying It\^{o}'s formula to $(h_t X^{\hat\pi}_t-a)^2$, we have
\begin{align*}
d(h_t X^{\hat\pi}_t-a)^2=(h_t X^{\hat\pi}_t-a)^2\Big[\frac{\xi_t^2}{Y_t^2}-2\frac{1}{Y_t}\xi_t'\phi_t\Big]dt-2(h_t X^{\hat\pi}_t-a)^2\frac{1}{Y_t}\xi_t'dW_t.
\end{align*}
Recalling that $Y$ follows \eqref{Y} and applying It\^{o}'s formula to $\frac{1}{Y_t}(h_t X^{\hat\pi}_t-a)^2$, we have
\begin{align*}
d\Big[\frac{1}{Y_t}(h_t X^{\hat\pi}_t-a)^2\Big]&=\frac{(h_t X^{\hat\pi}_t-a)^2}{Y_t}\Big[\frac{\xi_t^2}{Y_t^2}-2\frac{1}{Y_t}\xi_t'\phi_t\Big]dt\\
&\quad-\frac{(h_t X^{\hat\pi}_t-a)^2}{Y_t^3}\Big[\dist_{\sigma'\Gamma}^2(\phi Y-Z)-|\phi Y-Z|^2\Big]dt+2\frac{(h_t X^{\hat\pi}_t-a)^2}{Y_t^3}\xi_t'Zdt\\
&\quad-\frac{(h_t X^{\hat\pi}_t-a)^2}{Y_t^2}Z'dW_t-2\frac{(h_t X^{\hat\pi}_t-a)^2}{Y_t}\xi_t'dW_t\\
&=\frac{(h_t X^{\hat\pi}_t-a)^2}{Y_t^3}\Big[\xi_t^2-2\xi_t'\phi_tY_t-\dist_{\sigma'\Gamma}^2(\phi Y-Z)+|\phi Y-Z|^2+2\xi_t'Z\Big]dt\\
&\quad-\frac{(h_t X^{\hat\pi}_t-a)^2}{Y_t^2}(Z+2Y\xi)'dW_t\\
&=-\frac{(h_t X^{\hat\pi}_t-a)^2}{Y_t^2}(Z+2Y\xi)'dW_t.
\end{align*}
Since the stochastic integral in the last equation is a local martingale, there exists an increasing sequence of stopping times $\tau_n$ such that $\tau_n\uparrow+\infty$ as $n\rightarrow+\infty$ such that
\begin{align*}
\E\Big[\frac{1}{Y_{\iota\wedge\tau_n}}(h_{\iota\wedge\tau_n} X^{\hat\pi}_{\iota\wedge\tau_n}-a)^2\Big]=\frac{1}{Y_0}(h_0x-a)^2=\frac{Y_0}{\theta^2},
\end{align*}
for any stopping time $\iota\leq T$.
Letting $n\rightarrow\infty$, it follows from Fatou's lemma that
\begin{align*}
\E\Big[\frac{1}{Y_{\iota}}(h_{\iota} X^{\hat\pi}_{\iota}-a)^2\Big]\leq \frac{Y_0}{\theta^2}.
\end{align*}
Since $c_1\leq Y$, $h\leq c_2$ for some constants $c_2>c_1>0$, we get
\begin{align}\label{X2int}
\E\Big[ (X^{\hat\pi}_{\iota})^2\Big]\leq c_3,
\end{align}
for some constant $c_3$ and any stopping time $\iota\leq T$. Now it is standard to prove $\hat\pi\in L^2_{\mathcal{F}}(0,T;\mathbb{R}^m)$; see, e.g., \cite[Lemma 4.3]{HSX}. This proves $\hat\pi\in\Pi$.


By virtue of \eqref{LambdatoX} and \eqref{X2int}, $c_1\leq Y$, $h\leq c_2$, we immediately get that there exists some positive constant $c_4$,
\begin{align}\label{Lambdasqu}
\E[(\Lambda^{\hat\eta}_{\iota})^2]<c_4,
\end{align}
for any stopping time $\iota\leq T$. This proves $\hat\eta\in\mathcal{A}$.

\textbf{Step 2:} To prove \eqref{saddle1}.
For any $\pi\in\Pi$, applying It\^{o}'s formula to $R^{(\hat\eta,\pi)}=X_t^{\pi}h_t+\frac{1}{2\theta}(\Lambda_t^{\hat\eta} Y_t-1)$, we have
\begin{align*}
\E^{\mathbb{P}^{\hat\eta}}[R^{(\hat\eta,\pi)}_{t\wedge\tau_n}]&=xh_0+\frac{1}{2\theta}(Y_0-1)
+\E^{\mathbb{P}^{\hat\eta}}\int_0^{t\wedge\tau_n}\Big[h\pi'(\mu + \sigma\hat\eta)+\frac{1}{2\theta}\Lambda_t Y_t|\hat\eta|^2+\frac{1}{\theta}\Lambda \hat\eta'Z-\frac{1}{2\theta}\Lambda f\Big]ds.
\end{align*}

On one hand, for any $\pi\in\Pi$, setting $u=\sigma'\pi$ in Lemma \ref{proj}, we get
\begin{align}\label{substep1}
h\pi'(\mu + \sigma\hat\eta)=\frac{h}{Y}\pi'\sigma(\phi Y-Z-\xi )\leq \frac{h}{Y}\xi'(\phi Y-Z-\xi)=0,
\end{align}
where the last equality comes from \eqref{projpro}.

On the other hand, recall \eqref{funf} and \eqref{xi}:
\begin{align}\label{substep2}
Y|\hat\eta|^2+2 \hat\eta'Z- f&=Y|\hat\eta|^2+2 \hat\eta'Z+\frac{1}{Y}\dist^2_{\sigma'\Gamma}(\phi Y-Z)-\frac{1}{Y}|\phi Y-Z|^2+\frac{1}{ Y}|Z|^2\nn\\
&=Y\Big|\hat\eta+\frac{1}{Y}Z\Big|^2+\frac{1}{Y}\dist^2_{\sigma'\Gamma}(\phi Y-Z)-\frac{1}{Y}|\phi Y-Z|^2\nn\\
&=\frac{1}{Y}\Big[|\xi|^2+\dist^2_{\sigma'\Gamma}(\phi Y-Z)-|\phi Y-Z|^2\Big]\nn\\
&=\frac{1}{Y}\Big[|\xi|^2+\Big|\xi-(\phi Y-Z)\Big|^2-|\phi Y-Z|^2\Big]\nn\\
&=\frac{1}{Y}\Big[2|\xi|^2-2\xi'(\phi Y-Z)\Big]=0,
\end{align}
where we used \eqref{projpro} in the last equality.

Combining \eqref{substep1} and \eqref{substep2}, we arrive at
\begin{align}\label{step2}
\E^{\mathbb{P}^{\hat\eta}}[R^{(\hat\eta,\pi)}_{t\wedge\tau_n}]=\E[\Lambda^{\hat\eta}_{t\wedge\tau_n}R^{(\hat\eta,\pi)}_{t\wedge\tau_n}]\leq xh_0+\frac{1}{2\theta}(Y_0-1), \ \forall t\in[0,T], \ \forall\pi\in\Pi.
\end{align}

For any $\pi\in\Pi$, it is standard to verify $\E[\sup_{t\in[0,T]}(X^{\pi}_t)^2]<\infty$.
Since $h$ and $Y$ are bounded, we get
\begin{align*}
\E\Big[\sup_{t\in[0,T]}|\Lambda^{\hat\eta}_tR^{\hat\eta,\pi}_t|\Big]
&\leq C\E\Big[\sup_{t\in[0,T]}|\Lambda^{\hat\eta}_tX_t^{\pi}|\Big]+C\E\Big[\sup_{t\in[0,T]}(\Lambda^{\hat\eta}_t)^2\Big]+C\E\Big[\sup_{t\in[0,T]}\Lambda^{\hat\eta}_t\Big]\\
&\leq C\E\Big[\sup_{t\in[0,T]}(X^{\pi}_t)^2\Big] +C\E\Big[\sup_{t\in[0,T]}(\Lambda^{\hat\eta}_t)^2\Big]+C\\
&\leq C\E\Big[\sup_{t\in[0,T]}(X^{\pi}_t)^2\Big]+C\E\Big[(\Lambda_T^{\hat\eta})^2\Big]+C<\infty,
\end{align*}
where $C>0$ denotes a constant that may vary from line to line;
we used the elementary inequality $2ab\leq a^2+b^2$ in the second inequality, Doob's inequality in the third inequality, and \eqref{Lambdasqu} in the fourth inequality. Sending $n\rightarrow\infty$ in \eqref{step2} and using the dominated convergence theorem, we get \eqref{saddle1}.

\textbf{Step 3:} To prove \eqref{saddle2}.
For any $\eta\in\mathcal{A}$,  we have
\begin{align}
\E^{\mathbb{P}^{\eta}}[R^{(\eta,\hat\pi)}_{t\wedge\tau_n}]
&=xh_0+\frac{1}{2\theta}(Y_0-1)
+\E^{\mathbb{P}^{\eta}}\int_0^{t\wedge\tau_n}\frac{\Lambda^{\eta}}{2\theta}\Big[ Y_t|\eta|^2+2 \eta'Z+\frac{2\theta}{\Lambda^{\eta}}h\eta'\sigma'\hat\pi+\frac{2\theta}{\Lambda^{\eta}}h\hat\pi'\sigma\phi- f\Big]ds\nonumber\\
&=xh_0+\frac{1}{2\theta}(Y_0-1)
+\E^{\mathbb{P}^{\eta}}\int_0^{t\wedge\tau_n}\frac{\Lambda^{\eta}}{2\theta}\Big[Y\Big|\eta+\Big( \frac{Z}{Y}+\frac{\theta}{\Lambda^{\eta} Y}h\sigma'\hat\pi\Big)\Big|^2\nonumber
\\
&\qquad\qquad\qquad\qquad\qquad\qquad\qquad\qquad-\frac{1}{ Y}\Big| Z+\frac{\theta}{\Lambda^{\eta}}h\sigma'\hat\pi\Big|^2+\frac{2\theta}{\Lambda^{\eta}}h\hat\pi'\sigma\phi- f\Big]ds\nonumber\\
&\geq xh_0+\frac{1}{2\theta}(Y_0-1)
+\E^{\mathbb{P}^{\eta}}\int_0^{t\wedge\tau_n}\frac{\Lambda^{\eta}}{2\theta}\Big[-\frac{1}{ Y}\Big| Z+\frac{\theta}{\Lambda^{\eta}}h\sigma'\hat\pi\Big|^2+\frac{2\theta}{\Lambda^{\eta}}h\hat\pi'\sigma\phi- f\Big]ds.\label{goon}
\end{align}
Please note that $\hat\pi$ is a feedback of $\Lambda$, so under $\eta$, the optimal $\hat\pi$ in \eqref{hatpi} should take
\begin{align*}
\hat\pi =\frac{\Lambda^{\eta}} {h\theta}(\sigma\sigma')^{-1} \sigma\mbox{Proj}_{\sigma'\Gamma}(Y\phi-Z).
\end{align*}
Recall \eqref{hatsigmapi2}, and the definition of   $f$ in  \eqref{funf}, we have
\begin{align*}
&\quad-\frac{1}{ Y}\Big| Z+\frac{\theta}{\Lambda^{\eta}}h\sigma'\hat\pi\Big|^2+\frac{2\theta}{\Lambda^{\eta}}h\hat\pi'\sigma\phi- f\\
&=-\frac{1}{Y}\Big| Z+\mbox{Proj}_{\sigma'\Gamma}(Y\phi-Z)\Big|^2+2\phi'\mbox{Proj}_{\sigma'\Gamma}(Y\phi-Z)-f=0.
\end{align*}
Hence from \eqref{goon}, we have $$\E^{\mathbb{P}^{\eta}}[R^{(\eta,\hat\pi)}_{t\wedge\tau_n}]\geq xh_0+\frac{1}{2\theta}(Y_0-1).$$
Similar to Step 2, we can prove $\E[\sup_{t\in[0,T]}|\Lambda^\eta_tR_t^{\eta,\hat\pi}|]<\infty$. By sending $n\rightarrow\infty$ we get \eqref{saddle2}.

\textbf{Step 4:} To prove \eqref{saddle3}.
Applying It\^{o}'s formula to $R^{(\hat\eta,\hat\pi)}=X_t^{\hat\pi}h_t+\frac{1}{2\theta}(\Lambda_t^{\hat\eta} Y_t-1)$, we have
\begin{align*}
\E^{\mathbb{P}^{\hat\eta}}[R^{(\eta,\hat\pi)}_{t\wedge\tau_n}]
&=xh_0+\frac{1}{2\theta}(Y_0-1)
+\E^{\mathbb{P}^{\hat\eta}}\int_0^{t\wedge\tau_n}\Big[\frac{1}{2\theta}\Lambda_t Y_t|\hat\eta|^2+\frac{1}{\theta}\Lambda \hat\eta'Z+h\hat\eta'\sigma'\hat\pi+h\hat\pi'\mu-\frac{1}{2\theta}\Lambda f\Big]ds\\
&=xh_0+\frac{1}{2\theta}(Y_0-1)
+\E^{\mathbb{P}^{\hat\eta}}\int_0^{t\wedge\tau_n}\Big[\frac{\Lambda Y}{2\theta}\Big|\hat\eta+\frac{\theta}{\Lambda Y}\Big(\frac{1}{\theta}\Lambda Z+h\sigma'\hat\pi\Big)\Big|^2
\\
&\qquad\qquad\qquad\qquad\qquad\qquad-\frac{\theta}{2\Lambda Y}\Big|\frac{1}{\theta}\Lambda Z+h\sigma'\hat\pi\Big|^2+h\hat\pi'\mu-\frac{1}{2\theta}\Lambda f\Big]ds\\
&= xh_0+\frac{1}{2\theta}(Y_0-1),
\end{align*}
where the third equality comes from the definitions of $f$, $\hat\pi$, and $\hat\eta$ in \eqref{funf}, \eqref{hatpi}, and \eqref{saddleeta}.
Similar to Step 2, we can prove $\E[\sup_{t\in[0,T]}|\Lambda^\eta_tR_t^{\eta,\hat\pi}|]<\infty$ and \eqref{saddle3}.
\eof
%

As a byproduct of Theorem \ref{verif}, we have the following results for problem \eqref{MMV}.
\begin{corollary}
Let $(Y,Z)$ be the unique uniformly positive solution to \eqref{Y}, and let $h$, $\hat\pi$ be defined in \eqref{h} and \eqref{hatpi}, respectively. Then
$\hat\pi$ is optimal for the constrained MMV problem \eqref{MMV}, and
\begin{align}\label{MMVvalue}
\sup_{\pi\in\Pi}\inf_{\eta\in\mathcal{A}}\E^{\mathbb{P}^{\eta}}\Big[X_T+\frac{1}{2\theta}(\Lambda^\eta_T-1)\Big]
=xh_0+\frac{1}{2\theta}(Y_0-1).
\end{align}
\end{corollary}

\begin{remark}
If $\Gamma=\mathbb{R}^m$, then
\[\mbox{Proj}_{\sigma'\Gamma}(\phi Y-Z)
=\phi Y-\sigma'(\sigma\sigma')^{-1}\sigma Z.\]
Indeed, it suffices to prove
$$\mbox{Proj}_{\sigma'\Gamma}(\phi Y- Z)=\mbox{Proj}_{\sigma'\Gamma}(\phi Y-\sigma'(\sigma\sigma')^{-1}\sigma Z),$$
since $\phi Y-\sigma'(\sigma\sigma')^{-1}\sigma Z\in \sigma'\mathbb{R}^m=\sigma'\Gamma$.
Let $\xi=\mbox{Proj}_{\sigma'\Gamma}(\phi Y-Z)$ be defined by \eqref{xi}.
Then, for any $u\in\sigma'\Gamma$,
\begin{align*}
(u-\xi)'(\phi Y-\sigma'(\sigma\sigma')^{-1}\sigma Z-\xi)&=(u-\xi)'(\phi Y-Z+(I_n-\sigma'(\sigma\sigma')^{-1}\sigma) Z-\xi)\\
&=(u-\xi)'(\phi Y-Z-\xi)+(u-\xi)'(I_n-\sigma'(\sigma\sigma')^{-1}\sigma) Z\\
&\leq (u-\xi)'(I_n-\sigma'(\sigma\sigma')^{-1}\sigma) Z\\
&=0,
\end{align*}
where we used Lemma \ref{proj} to get the inequality and
the fact $u-\xi\in\sigma'\Gamma$ and $\sigma(I_n-\sigma'(\sigma\sigma')^{-1}\sigma)=0$ to get the last equality. Thanks to Lemma \ref{proj} again, this inequality implies $\xi= \mbox{Proj}_{\sigma'\Gamma}(\phi Y-\sigma'(\sigma\sigma')^{-1}\sigma Z)$.

Therefore, by \eqref{funf}, \eqref{hatpi}, and \eqref{saddleeta},
\begin{align*}
f &=\frac{1}{Y}(\phi Y-\sigma Z)'\sigma'(\sigma\sigma')^{-1}\sigma(\phi Y- Z)-\frac{1}{ Y}|Z|^2,\\
\hat \pi &=\frac{\Lambda}{\theta h}(\sigma\sigma')^{-1}\sigma(\phi Y- Z),\\
\hat\eta&=-\frac{1}{Y}(Z+\sigma'(\sigma\sigma')^{-1}\sigma(\phi Y- Z))=-\sigma'(\sigma\sigma')^{-1}\sigma \phi-(I_n-\sigma'(\sigma\sigma')^{-1}\sigma)\frac{Z}{Y}.
\end{align*}
If $m=n$, then $\sigma$ is invertible, and $\hat\eta=-\phi=-\sigma^{-1}\mu$.
\end{remark}

\section{Comparison with the classical MV problem}\label{Comparison}
The constrained classical MV problem is (see, e.g., \cite{ZL}) is
\begin{align}\label{MV}
\sup_{\pi\in\Pi}\Big[\E(X_T)-\frac{\theta}{2}\mbox{Var}(X_T)\Big].
\end{align}
We will firstly solve the problem \eqref{MV} and then draw a comparison with the MMV problem \eqref{MMV}.

For any $K\in\mathbb{R}$, let us consider
\begin{align}\label{MV1}
F(K):=\inf_{\pi\in\Pi^K}\E\Big[(X_T-K)^2\Big]=\inf_{\pi\in\Pi^K}\Big[\E(X_T^2)-K^2\Big],
\end{align}
where
\begin{align*}
\Pi^K:=\Big\{\pi\in\Pi\;\Big|\;\E(X_T^\pi)=K\Big\}
\end{align*}
with the convention of $F(K)=+\infty$ if $\Pi^K=\emptyset$. Trivially, $F(K)\geq0$.

As illustrated in \cite[page 12]{TZ},
the connection between problems \eqref{MV} and \eqref{MV1} is given as follows:
\begin{align}
\sup_{\pi\in\Pi}\Big[\E(X_T)-\frac{\theta}{2}\mbox{Var}(X_T)\Big]
&=\sup_{K\in\mathbb{R}}\sup_{\pi\in\Pi^K}\Big[K-\frac{\theta}{2}\Big(E(X_T^2)-K^2\Big)\Big]\nn\\
&=\sup_{K\in\mathbb{R}}\Big[K-\frac{\theta}{2}\inf_{\pi\in\Pi^K}\Big(E(X_T^2)-K^2\Big)\Big]
=\sup_{K\in\mathbb{R}}\Big[K-\frac{\theta}{2} F(K)\Big].\label{Lag0}
\end{align}

In order to solve \eqref{MV1}, we introduce a Lagrange multiplier $\gamma\in\mathbb{R}$ and consider
\begin{align}\label{MV2}
J(K,\gamma):=\inf_{\pi\in\Pi}\E\Big[{X_T^2-K^2-2\gamma(X_T-K)}\Big]
=\inf_{\pi\in\Pi}\Big[\E(X_T-\gamma)^2-(K-\gamma)^2\Big].
\end{align}
By the Lagrange duality theorem (see Luenberger \cite{Lu}),
\begin{align}\label{Lag1}
F(K)=\inf_{\pi\in\Pi^K}\Big(\E(X_T^2)-K^2\Big)&=\sup_{\gamma\in\mathbb{R}}J(K,\gamma), \ K\in\mathbb{R}.
\end{align}

The solution for problem \eqref{MV2} depends on the following two BSDEs:
\begin{align}
\label{P1}
\begin{cases}
dP_1=-\Big\{2rP_1+\inf\limits_{\pi\in\Gamma}\Big[P_1\pi'\sigma\sigma'\pi+2\pi'(P_1\mu+\sigma\Delta_1)\Big]\Big\}dt+\Delta_1'dW, \\
P_{1,T}=1, \quad
P_{1,t}>0, \ \mbox{ for $t\in[0, T]$,}
\end{cases}
\end{align}
and
\begin{align}
\label{P2}
\begin{cases}
dP_2=-\Big\{2rP_2+\inf\limits_{\pi\in\Gamma}\Big[P_2\pi'\sigma\sigma'\pi-2\pi'(P_2\mu+\sigma\Delta_2)\Big]\Big\}dt+\Delta_2'dW, \\
P_{2,T}=1, \quad P_{2,t}>0, \ \mbox{ for $t\in[0, T]$.}
\end{cases}
\end{align}
Recall that $h_t=e^{\int_t^Tr_sds}$ is deterministic and bounded.
From \cite[Theorem 4.2 and Theorem 5.2]{HZ}, there exists a unique uniformly positive solution $(P_1,\Delta_1)\in L^{\infty}_{\mathcal{F}}(0,T;\mathbb{R})\times L^{2}_{\mathcal{F}}(0,T;\mathbb{R}^n)$ to the BSDE \eqref{P1}. And
$P_{1,0}\leq h_0^2$ from \cite[Lemma 6.1]{HZ}. Similar results hold for the BSDE \eqref{P2}
with $P_{2,0}\leq h_0^2$.

The following results for the problem \eqref{MV2} generalize a little bit \cite[Theorem 6.2]{HZ} from $\Gamma=\R^m_+$ to a general closed convex cone $\Gamma\subseteq\R^m$ with exactly the same proof.
\begin{lemma}
Let $(P_1,\Delta_1)$ and $(P_2,\Delta_2)$ be the unique uniformly positive solutions to \eqref{P1} and \eqref{P2}, respectively. Then the feedback strategy
\begin{align}\label{pistar}
\pi^\gamma(t, X)=\Big(X-\frac{\gamma}{h_t}\Big)^+ \xi_1 +\Big(X-\frac{\gamma} {h_t}\Big)^-\xi_2,
\end{align}
is optimal for the problem \eqref{MV2}, where
\begin{align}\label{xi12}
\xi_1:=(\sigma\sigma')^{-1}\sigma Proj_{\sigma'\Gamma}\Big(-\phi-\frac{\Delta_1}{P_1}\Big) \quad \mbox{and} \quad \xi_2:=(\sigma\sigma')^{-1}\sigma Proj_{\sigma'\Gamma}\Big(\phi+\frac{\Delta_2}{P_2}\Big).
\end{align}
And the optimal value of \eqref{MV2} is
\begin{align*}
J(K,\gamma)=
\begin{cases}
J_1(K,\gamma), \ \mbox{if} \ \gamma< x h_0;\\
J_2(K,\gamma), \ \mbox{if} \ \gamma\geq x h_0,
\end{cases}
\end{align*}
where
\begin{align*}
J_1(K,\gamma) &:=\Big(\frac{P_{1,0}}{h_0^2}-1\Big)\gamma^2-2\Big(x\frac{P_{1,0}}{h_0}-K\Big)\gamma+P_{1,0}x^2-K^2, \\ 
J_2(K,\gamma) &:=\Big(\frac{P_{2,0}}{h_0^2}-1\Big)\gamma^2-2\Big(x\frac{P_{2,0}}{h_0}-K\Big)\gamma+P_{2,0}x^2-K^2.
\end{align*}
\end{lemma}

To solve \eqref{Lag1},
we do some tedious calculation and obtain
\begin{align*}
\sup_{\gamma<xh_0}J_1(K,\gamma)
=
\begin{cases}
-(K-xh_0)^2, & \ \mbox{if} \ K> xh_0 ;\\
+\infty, & \ \mbox{if} \ K<xh_0 \ \mbox{and} \ P_{1,0}=h_0^2;\\
\frac{P_{1,0}(K-xh_0)^2}{h_0^2-P_{1,0}}, & \ \mbox{if} \ K<xh_0 \ \mbox{and} \ P_{1,0}<h_0^2;\\
0, & \ \mbox{if} \ K=xh_0,
\end{cases}
\end{align*}
and
\begin{align*}
\sup_{\gamma\geq xh_0}J_2(K,\gamma)
=
\begin{cases}
+\infty, & \ \mbox{if} \ K>xh_0 \ \mbox{and} \ P_{2,0}=h_0^2;\\
\frac{P_{2,0}(K-xh_0)^2}{h_0^2-P_{2,0}}, & \ \mbox{if} \ K>xh_0 \ \mbox{and} \ P_{2,0}<h_0^2;\\
-(K-xh_0)^2, & \ \mbox{if} \ K< xh_0 ;\\
0, & \ \mbox{if} \ K=xh_0.
\end{cases}
\end{align*}
Therefore,
\begin{align*}
F(K)=\sup_{\gamma\in\mathbb{R}}J(K,\gamma)&=\max\Big\{\sup_{\gamma< xh_0}J_1(K,\gamma),\sup_{\gamma\geq xh_0}J_2(K,\gamma)\Big\}\\
&=
\begin{cases}
+\infty, & \ \mbox{if} \ K> xh_0 \ \mbox{and} \ P_{2,0}=h_0^2;\\
\frac{P_{2,0}(K-xh_0)^2}{h_0^2-P_{2,0}}, & \ \mbox{if} \ K> xh_0 \ \mbox{and} \ P_{2,0}<h_0^2;\\
+\infty, & \ \mbox{if} \ K<xh_0 \ \mbox{and} \ P_{1,0}=h_0^2;\\
\frac{P_{1,0}(K-xh_0)^2}{h_0^2-P_{1,0}}, & \ \mbox{if} \ K<xh_0 \ \mbox{and} \ P_{1,0}<h_0^2;\\
0, & \ \mbox{if} \ K=xh_0,
\end{cases}
\end{align*}
with the argument maximum
\begin{align}\label{hatgammaK}
\hat\gamma(K)=
\begin{cases}
+\infty, & \ \mbox{if} \ K> xh_0 \ \mbox{and} \ P_{2,0}=h_0^2;\\
\frac{h_0^2K-xP_{2,0}h_0}{h_0^2-P_{2,0}}, & \ \mbox{if} \ K> xh_0 \ \mbox{and} \ P_{2,0}<h_0^2;\\
-\infty, & \ \mbox{if} \ K<xh_0 \ \mbox{and} \ P_{1,0}=h_0^2;\\
\frac{h_0^2K-xP_{1,0}h_0}{h_0^2-P_{1,0}}, & \ \mbox{if} \ K<xh_0 \ \mbox{and} \ P_{1,0}<h_0^2;\\
xh_0, & \ \mbox{if} \ K=xh_0.
\end{cases}
\end{align}
Accordingly,
\begin{align*}
K-\frac{\theta}{2}F(K)=
\begin{cases}
-\infty, & \ \mbox{if} \ K> xh_0 \ \mbox{and} \ P_{2,0}=h_0^2;\\
K-\frac{\theta}{2}\frac{P_{2,0}(K-xh_0)^2}{h_0^2-P_{2,0}}, & \ \mbox{if} \ K> xh_0 \ \mbox{and} \ P_{2,0}<h_0^2;\\
-\infty, & \ \mbox{if} \ K<xh_0 \ \mbox{and} \ P_{1,0}=h_0^2;\\
K-\frac{\theta}{2}\frac{P_{1,0}(K-xh_0)^2}{h_0^2-P_{1,0}}, & \ \mbox{if} \ K<xh_0 \ \mbox{and} \ P_{1,0}<h_0^2;\\
xh_0, & \ \mbox{if} \ K=xh_0.
\end{cases}
\end{align*}
And
\begin{align}\label{MVvalue2}
\sup_{K\in\mathbb{R}}[K-\frac{\theta}{2}F(K)]&=
\begin{cases}
xh_0+\frac{1}{2\theta}\Big(\frac{h_0^2}{P_{2,0}}-1\Big), & \ \mbox{if} \ P_{2,0}<h_0^2;\\
xh_0, & \ \mbox{if} \ P_{2,0}=h_0^2,
\end{cases}\nn\\
&=xh_0+\frac{1}{2\theta}\Big(\frac{h_0^2}{P_{2,0}}-1\Big),
\end{align}
with the argument maximum
\begin{align}\label{hatK}
\hat K= x h_0+\frac{1}{\theta}\Big(\frac{h_0^2}{P_{2,0}}-1\Big)\geq xh_0.
\end{align}
Substituting \eqref{hatK} into \eqref{hatgammaK}, we obtain
\begin{align*}
\hat\gamma(\hat K)=xh_0+\frac{1}{\theta}\frac{h_0^2}{P_{2,0}}.
\end{align*}
%

The above analysis leads to the following results for the problem \eqref{MV}.
\begin{theorem}
Let $\xi_2$ be defined in \eqref{xi12}, and let $(P_1,\Delta_1)$ and $(P_2,\Delta_2)$ be the unique uniformly positive solutions to \eqref{P1} and \eqref{P2}, respectively.
Let
\begin{align}
\hat\gamma=xh_0+\frac{1}{\theta}\frac{h_0^2}{P_{2,0}},
\end{align}
and
\begin{align}\label{hatpimv}
\pi^{\hat\gamma}(t,X)=-\Big( X-\frac{\hat\gamma}{h_t}\Big)\xi_2.
\end{align}
Then $\pi^{\hat\gamma}$ is the optimal feedback portfolio for the problem \eqref{MV}, and
\begin{align}\label{MVvalue}
\sup_{\pi\in\Pi}\Big[\E(X_T)-\frac{\theta}{2}\mbox{Var}(X_T)\Big]
=xh_0+\frac{1}{2\theta}\Big(\frac{h_0^2}{P_{2,0}}-1\Big).
\end{align}
\end{theorem}
\pf
Notice $\hat\gamma=\hat\gamma(\hat K)$; so, by \eqref{Lag0} and \eqref{Lag1},
\begin{align*}
\sup_{\pi\in\Pi}\Big[\E(X_T)-\frac{\theta}{2}\mbox{Var}(X_T)\Big]
&=\sup_{K\in\mathbb{R}}\Big[K-\frac{\theta}{2} F(K)\Big)\Big]\\
&=\hat K-\frac{\theta}{2} F(\hat K)
=\hat K-\frac{\theta}{2} J(\hat K,\hat\gamma(\hat K))
=\hat K-\frac{\theta}{2} J(\hat K,\hat\gamma).
\end{align*}
Therefore, the optimal portfolio for the problem \eqref{MV2} with $(K,\gamma)=(\hat K,\hat\gamma)$ is also optimal to \eqref{MV}.

We now show that the portfolio \eqref{hatpimv} is optimal to the problem \eqref{MV2} with $(K,\gamma)=(\hat K,\hat\gamma)$.
Taking \eqref{hatpimv} to the wealth process \eqref{wealth}, we get
\begin{align*}
d\Big(X_t^{\pi^{\hat\gamma}}-\frac{\hat\gamma}{h_t}\Big)=\Big(X_t^{\pi^{\hat\gamma}}-\frac{\hat\gamma}{h_t}\Big)
\Big((r-\xi_2'\mu)dt-\xi_2\sigma'dW\Big),
\end{align*}
then It\^{o}'s lemma gives
\begin{align*}
X_t^{\pi^{\hat\gamma}}-\frac{\hat\gamma}{h_t}=\Big(X^{\pi^{\hat\gamma}}_0-\frac{\hat\gamma}{h_0}\Big)
\exp\Big(\int_0^t(r-\xi_2'\mu-\frac{1}{2}|\xi_2'\sigma|^2)ds-\int_0^t\xi_2'\sigma dW_s\Big).
\end{align*}
Since $X^{\pi^{\hat\gamma}}_0-\frac{\hat\gamma}{h_0}=-\frac{h_0}{\theta P_{2,0}}\leq 0$, it follows that $X_t^{\pi^{\hat\gamma}}-\frac{\hat\gamma}{h_t}\leq 0$.
Accordingly, the portfolio \eqref{hatpimv} is
\begin{align*}
\pi^{\hat\gamma}(t, X)=-\Big( X-\frac{\hat\gamma}{h_t}\Big)\xi_2=\Big(X-\frac{\hat\gamma}{h_t}\Big)^+ \xi_1 +\Big(X-\frac{\hat\gamma} {h_t}\Big)^-\xi_2,
\end{align*}
which is just \eqref{pistar}, the optimal strategy for the problem \eqref{MV2}.

Finally, \eqref{MVvalue} comes from \eqref{Lag0} and \eqref{MVvalue2} evidently.
\eof

In the end, we have the following connection between problems \eqref{MMV} and \eqref{MV}.
\begin{theorem}\label{com}
The problems \eqref{MMV} and \eqref{MV} have the same optimal value $$xh_0+\frac{1}{2\theta}(Y_0-1)$$ and optimal feedback portfolio
\begin{align*}
\hat\pi=\pi^{\hat\gamma},
\end{align*}
where $\hat\pi$ and $\pi^{\hat\gamma}$ are defined in \eqref{hatpi} and
\eqref{hatpimv}, respectively.
\end{theorem}
\pf
Recall that $(Y,Z)$ is the unique uniformly positive solution to \eqref{Y}, and $h_t=e^{\int_t^Tr_sds}$. It can be directly verified, by It\^{o}'s formula, that
\begin{align}\label{YtoP}
(Y,Z)=\Big(\frac{h^2}{P_2},-\frac{h^2}{P^2_2}\Delta_2\Big).
\end{align}
Comparing \eqref{MMVvalue} and \eqref{MVvalue}, the first claim follows.

By \eqref{hatsigmapi2} and \eqref{YtoP},
\begin{align*}
\hat\pi&=\frac{xh_0+\frac{Y_0}{\theta}- h_t X^{\hat\pi}_t}{ h_tY_t}(\sigma\sigma')^{-1}\sigma \mbox{Proj}_{\sigma'\Gamma}(\phi Y-Z)\\
&=\frac{xh_0+\frac{h_0^2}{\theta P_{2,0}}- h_t X^{\hat\pi}_t}{ h_t^3}P_{2}(\sigma\sigma')^{-1}\sigma \mbox{Proj}_{\sigma'\Gamma}\Big(\phi \frac{h^2}{P_2}+\frac{h^2}{P^2_2}\Delta_2\Big)\\
&=\frac{\hat\gamma- h_t X^{\hat\pi}_t}{ h_t}(\sigma\sigma')^{-1}\sigma \mbox{Proj}_{\sigma'\Gamma}\Big(\phi +\frac{\Delta_2}{P_2}\Big)\\
&=-\Big( X_t^{\hat\pi}-\frac{\hat\gamma}{h_t}\Big)\xi_2,
\end{align*}
which is exactly the feedback optimal portfolio \eqref{hatpimv}.
\eof

\section{Concluding remarks}\label{conclude}
In this paper, we study the MMV and the MV problems with cone constraints on the portfolios in a diffusion model with random coefficients. Semiclosed solutions for both problems in terms of the solutions to some BSDEs are provided. And by a careful comparison, we find that the solutions to these two problems are exactly the same. Further research along this line can be interesting as well; for instance: (1)
When $r$ is random, how does one solve the problem \eqref{MMV} even without any constraints on the portfolio? (2) When the wealth dynamics are discontinuous (e.g., with Poisson jumps), how does one solve the problem \eqref{MMV}?

Both Du and Strub \cite{DS23} and Li, Liang, and Pang \cite{LLP23} claim that
the optimal values and optimal strategies corresponding to the MMV and the MV problems coincide. The former paper assumes that the asset prices are continuous and that not investing in the risky assets is an admissible strategy, whereas the latter considers a jump diffusion model with stochastic factor but no trading constraint.
It is still an open question that whether the consistency still holds in a general model with discontinuous asset price or trading constraint.
\bigskip\\
\textbf{Acknowledgments.}
The authors thank the associated editor and anonymous referee for their valuable comments and suggestions. 
We thank the anonymous referee for pointing out the work \cite{LLP23} to us.
We also thank Strub for letting us know of his latest research work \cite{DS23}.

\end{document}